# ABSORBED DOSE IN ION BEAMS: COMPARISON OF IONISATION- AND FLUENCE-BASED MEASUREMENTS

Julia-Maria Osinga[1,2,*], Stephan Brons[3], James A. Bartz[4,5], Mark S. Akselrod[5], Oliver Jäkel[1,2,3], Steffen Greilich[2]

[1]Department of Radiation Oncology and Radiation Therapy, Heidelberg University Hospital, Im Neuenheimer Feld 400, 69120 Heidelberg, Germany
[2]Division of Medical Physics in Radiooncology, German Cancer Research Center (dkfz), Im Neuenheimer Feld 280, 69120 Heidelberg, Germany
[3]Heidelberg Ion-Beam Therapy Center, Im Neuenheimer Feld 450, 69120 Heidelberg, Germany
[4]Oklahoma State University, Stillwater, OK 74074, USA
[5]Landauer Inc., 723 1/2 Eastgate, Stillwater, OK 74074, USA

**Abstract**
A direct comparison measurement of fluorescent nuclear track detectors (FNTDs) and a thimble ionisation chamber is presented. Irradiations were performed using monoenergetic protons (142.66 MeV, $\phi = 3 \times 10^6$ 1/cm$^2$) and carbon ions (270.55 MeV/u, $\phi = 3 \times 10^6$ 1/cm$^2$). It was found that absorbed dose to water values as determined by fluence measurements using FNTDs are, in case of protons, in good agreement (2.4 %) with ionisation chamber measurements, if slower protons and Helium secondaries were accounted for by an effective stopping power. For carbon, however, a significant discrepancy of 4.5 % was seen, which could not be explained by fragmentation, uncertainties or experimental design. The results rather suggest a W-value of 32.10 eV ± 2.6 %. Additionally, the abundance of secondary protons expected from Monte-Carlo transport simulation was not observed.

*Corresponding author: e-mail: j.osinga@dkfz-heidelberg.de,
phone: +49-(0)6221-422633, fax: +49-(0)6221-422665

INTRODUCTION

Fluorescent nuclear track detectors (FNTDs) based on $Al_2O_3$:C,Mg single crystals and laser-scanning confocal fluorescence microscopy [1] allow for high-accuracy fluence determination. FNTDs exhibit excellent particle detection efficiency and can register all types of primary and secondary ions present in clinical beams (exemplary shown in Fig. 1, inserts) [2]. Potential applications of the FNTD technique are seen where employment of ionisation chambers is challenging, such as in laser-accelerated protons, dosimetry in magnetic fields or *in vivo* dosimetry. However, using FNTDs, discrepancies of ~8 % to ionisation-based measurements were observed in the authors´ studies. At the time, the findings were not conclusive owing to shortcomings in the experimental designs. In this contribution, a direct comparison study of FNTDs and a thimble ionisation chamber is presented to investigate this discrepancy in more detail.

MATERIALS AND METHODS

**Fluorescent nuclear track detectors**

$Al_2O_3$:C,Mg single crystals grown by Landauer Inc.,
Stillwater, OK, USA, were used as FNTDs (4x8x0.5 mm³ in size). $Al_2O_3$:C,Mg contains $F_2^{2+}$(2Mg) colour centres, which undergo radiochromic transformation under ionizing radiation yielding intra-centre fluorescence at 750±50 nm when stimulated at 620±50 nm. Since transformed centres are optically, thermally, and temporally stable, this enables optical imaging of energy deposition and hence charged particle tracks in three dimensions [3]. Further, it has been shown that the fluorescence amplitude of the particle tracks is related to the linear energy transfer (LET) of the particles enabling particle discrimination on a wide range of LET [4]. However, the performance of FNTDs for particle spectroscopy in clinical applications with the read-out protocol used within this study has still to be specified in more detail [5].

**Zeiss LSM 710 ConfoCor 3**

The Zeiss LSM 710 ConfoCor 3 inverted laser scanning confocal microscope was used for detector read-out with the configuration described in Ref. [6] [633 nm for excitation, 655 nm long-pass emission filter for detection, 63x/1.40NA oil-immersion objective lens with lateral (axial) resolution of ~200 nm (800 nm)].

**Image processing software**

ImageJ ([7], [8]) was used together with the 'Mosaic' background subtractor [9] and particle tracker [10] plug-ins for subtracting the fluorescence background and finding the particle track positions [2]. Further data processing was done in R (version 2.14.2) [11].

**Fluence-based dose approximation**

The absorbed dose to water for the beam quality $Q$ (e.g. p or $^{12}$C), $D_{w,Q}$, can be determined by the particle fluence, $\phi$, and the mass stopping-power of water, $\frac{s_{w,Q}}{\rho_w}$, through

$$D_{w,Q} = \phi \cdot \frac{s_{w,Q}}{\rho_w}. \qquad (1)$$

In case of mixed particle fields, dose contributions from different particle species $T$ and kinetic energies $E$ have to be considered:

$$D_{w,Q} = \frac{1}{\rho_w} \cdot \sum_T \int_E dE \cdot \phi_{E,T}(E) \cdot s_{w,Q}(E,T). \qquad (2)$$

In clinical ion beams, one can refer to the primary beam (index *prim*) and slower particles of the same type $T$ as well as secondaries due to scattering and nuclear fragmentation:

$$D_{w,Q} = \frac{1}{\rho_w} \cdot \begin{bmatrix} \phi_{prim}(E_{prim}) \cdot s_{w,Q}(E_{prim}) + \\ \int_0^{E<E_{prim}} dE \cdot \phi_{E,prim}(E) \cdot s_{w,Q}(E) + \\ \sum_{T \neq prim} \int_E dE \cdot \phi_{E,T}(E) \cdot s_{w,Q}(E,T) \end{bmatrix}. \qquad (3)$$

**Fluence assessment using FNTDs**

Within this study, the approach described in Ref. [2] was used to determine $\phi$, i.e. by

$$\phi = \frac{N}{A} \qquad (4)$$

where $N$ is the number of particles counted and $A$ the analysed area. In case of ions traversing the FNTD under a polar angle $\vartheta \neq 0°$ (*e.g.* non-perpendicular irradiation or misalignment of the FNTD under the microscope), $A$ is not the planar area $A_\perp$. This effect has been accounted for by multiplying $A$ with a correction factor, $k_A$,

$$A_\perp = k_A \cdot A = \cos\vartheta \cdot A \qquad (5)$$

with $\vartheta$ derived from the 3-d track structure information obtained within the FNTD. With respect to the carbon ion irradiations, particles were discriminated concerning the relative fluorescence amplitude of their tracks into primary particles and secondary lighter fragments in general. In case of the proton irradiations, no particle discrimination was applied (Figure 1, inserts). Since FNTDs have a track detection efficiency of ≥ 99.83 % [2] and uncertainties of $A$ have been proven to be negligible, the fluence uncertainty is dominated by (Poisson) counting statistics:

$$\sigma_{FNTD} = \frac{\Delta \phi}{\phi} = \frac{1}{\sqrt{N}} = \frac{1}{\sqrt{\phi \cdot A_\perp}}. \qquad (6)$$

The mass stopping-power values of water were taken from the ICRU reports 49 and 73 ([12], [13]).

**Ionisation chamber and ionisation-based dose**

A Farmer-type air-filled ionisation chamber was employed (0.6 cm³ sensitive volume, graphite-coated PMMA wall, PTW 30013). The absorbed dose to water was determined using the international code of practise TRS-398 [14]:

$$D_{w,Q} = M_Q \cdot N_{D,w,Q_0} \cdot k_{Q,Q_0} \qquad (7)$$

$$with\ k_{Q,Q_0} = \frac{(s_{w,air})_Q}{(s_{w,air})_{Q_0}} \cdot \frac{(W_{air})_Q}{(W_{air})_{Q_0}} \cdot \frac{p_Q}{p_{Q_0}} \qquad (8)$$

where $M_Q$ is the reading of the dosemeter (corrected for temperature and pressure, electrometer calibration, polarity and recombination) and $N_{D,w,Q_0}$ the calibration at reference quality $Q_0$ (here $^{60}$Co). $k_{Q,Q_0}$ (here 1.030) corrects for differences between the reference beam quality $Q_0$ and the actual beam quality $Q$ and relies on the stopping-power ratio (water to air), the $W$ value and the chamber-specific perturbation factor $p$. Before the first measurement, the correction factor $k_p/k_m$ was determined using a radioactive check device.

**Phantoms**

A water-equivalent RW-3 adaption plate (PTW, Freiburg, Germany) was used for the ionisation chamber (30 cm x 30 cm,

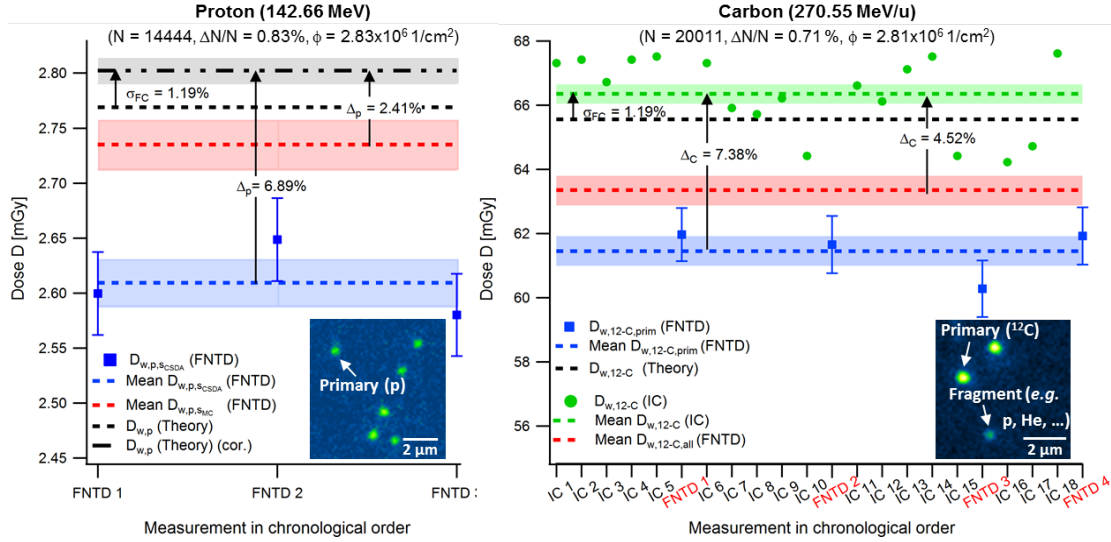

Figure 1: Results for the proton (left) and carbon (right) irradiations. $N$ is the total number of counted particle tracks with $\Delta N$ being the Poisson error and $\phi$ the corresponding mean particle fluence. The theoretical dose value $D_{w,Q}(Theory)$ (black dashed line) is obtained by multiplying the nominal particle fluence of $3 \times 10^6$ 1/cm$^2$ with $s_{w,Q}(E_{prim})$ (Table 1), whereas the dose assessed with FNTDs, $D(FNTD)$, was obtained by multiplying the measured particle fluence with the corresponding stopping power. The $D_{w,Q}(Theory)$ value was adjusted by Farmer chamber measurements (green line) performed during the carbon irradiations to $D_{w,Q}(Theory,cor.)$. Shadowed areas refer to the error of the mean. Inserts show how the corresponding particle tracks look like on the FNTD emphasising that they clearly stand out (even the low-LET protons) against the background. The colour scales were adapted to the specific images to allow for optimal contrast and are thus not comparable.

7 mm RW-3 in front and 10 mm RW-3 for backscatter). For the FNTD, 4.7 mm instead of 7 mm RW-3 were placed in front to obtain a compatible experimental set-up considering the effective point of measurement of the cylindrical ionisation chamber as given in Ref. [14]. For further calculations, a water-equivalent pathlength (WEPL) of $1.025 \pm 0.011$ was used for RW-3 [15].

**Particle energy and spectra**

Monte-Carlo (MC) transport simulations yielded information on the absorbed dose to water and particle fluences as a function of energy for primary and secondary particles. The FLUKA code ([16], [17]), version 2011 v2.17 was used. Scoring was done for a water volume (1x1x0.003 cm³) behind 7.7 mm of water. To study the potential influence of the phantom and the detector material, additional simulations were done where the water surrounding the target volume was replaced by RW-3 of corresponding thickness and the target volume by $Al_2O_3$, respectively.

EXPERIMENTS

**Irradiations**

Irradiations were performed at the Heidelberg Ion-Beam Therapy Center (HIT) with a field size of 10x10 cm$^2$. The phantoms were located at the iso-centre and irradiated with protons (142.66 MeV) and carbon ions (270.55 MeV/u) at a nominal fluence of $3 \times 10^6$ 1/cm$^2$. No ripple filter to broaden the Bragg-Peak was used. The beam application monitor system (BAMS) at HIT, featuring three ionisation chamber monitors, is calibrated in terms of particle fluence by a Farmer-type air-filled ionisation chamber on a daily basis allowing for a tolerance of $\pm 1$ %. To increase significance in this study, 18 additional measurements with the Farmer chamber placed in the RW-3 phantom were performed (carbon ions) and used to fine-tune the monitor chambers. Since this effect is independent of ion type, the adjustment was applied to the proton data as well. In total, four FNTDs (three FNTDs) were irradiated with carbon ions (protons).

**FNTD read-out**

All FNTDs were read-out 20 µm below the detector surface. "Z-stacks" of five images separated by $\Delta z = 1$ µm ($^1$H) and 5 µm ($^{12}$C) covering an area of 1.02 mm$^2$ were acquired. In order to improve the signal-to-noise ratio, a median intensity projection of the z-stacks was calculated where applicable [2].

**Irradiation-field homogeneity**

Physical beam records from the accelerator log system were forward-calculated and analyzed regarding deviations from the nominal particle fluence. Additionally, cross sections of the irradiated FNTDs in horizontal and vertical direction were acquired yielding a good approximation of the spatial fluence distribution.

Table 1: Monte-Carlo transport simulation results on CSDA energy, dose to water, and effective stopping power at 7.7 mm water-equivalent thicknesses (WET). Following WET were considered for the calculation of the particle energy at the detector surface ($E_{prim}$) using the CSDA by the "libamtrack" library [18]: (1) 2.89 mm, which includes all traversed materials between the high-energy beam line and the iso-centre, (2) 4.82 mm (4.7 mm RW-3). Stopping-powers for protons were taken from Ref. [12] and scaled for carbon ions using the effective ion charge.

| Primary | $E_{prim}$(CSDA) | $s_{CSDA}$ [keV/µm] | $E_{prim}$(MC) | $D_{water}$ [Gy·cm$^2$] | $\Phi_{rel}$ | $s_{MC}$ [keV/µm] | $\Delta s / s_{CSDA}$ |
|---|---|---|---|---|---|---|---|
| Proton | 138.29 | 0.5760 | 138.33±0.13 | 9.678×10$^{-10}$±0.11% | 1.004 | 0.6038 | +4.8 % |
| Carbon | 261.88 | 13.64 | 262.00±0.13 | 2.202×10$^{-8}$±0.11% | 1.183 | 13.69 | +0.3 % |

**Table 2: MC transport simulation results on relative fluences and doses for a water volume at WET of 7.7 mm.**

| Primary | Quantity | H Low E | H High E | He | Li | Be | B | C Low E | C High E |
|---|---|---|---|---|---|---|---|---|---|
| Proton | Fluence | 1.2 % | 98.8 % | <1 ‰ | <0.2 ‰ | - | - | - | - |
|  | Dose[1] | 4.0 % | 95.1 % | 0.7 % | 0.1 % | - | - | - | - |
| Carbon | Fluence | 14.8 % |  | 2.4 % | 0.3 % | 0.2 % | 0.4 % | 0.1 % | 81.8 % |
|  | Dose |  | 1.7 % | 0.6 % | 0.1 % | 0.1 % | 0.3 % | 0.2 % | 97.1 % |

[1] Additional 0.1 % relative dose from oxygen (O).

## RESULTS

The mean of the 18 Farmer chamber measurements performed in the carbon ion beam yielded an adjustment of $\sigma_{FC}$ = 1.19 % ± 0.01 pp (percentage point, SE) for the monitor system, i.e. the "corrected theoretical dose ($D_{w,Q}$(Theory,cor.))" (Figure 1).

### Protons

For protons, a deviation of $\Delta_p$ = 6.89 % between the mean fluence-based dose to water value of 2.61 mGy ± 0.83 % as obtained with FNTDs (Figure 1, left, blue line) and the ionisation-based value of 2.80 mGy ± 0.41 % (black line) was found assuming a monoenergetic proton beam with energy $E_{prim}$ = 138.3 MeV by a continuous-slowing-down approximation (CSDA) (Table 1). However, as indicated by the simulations, this is only true for $\Phi_{prim}(E_{prim})$ = 98.8 % of the protons detected, the remaining $\Phi_{prim}(E < E_{prim})$ = 1.2 % of lower-energy protons deposit a significant relative dose (4.0 %, Table 2), even in the entrance channel. Additionally, fragments like helium or lithium are very rare but still have a considerable contribution to dose due to their high stopping power. Taking these contributions into account by an effective stopping power (Table 1), the discrepancy $\Delta_p$ decreases to 2.4 %.

### Carbon ions

In case of carbon ions, both the primaries' fluence $\Phi_{prim}$ and the fluence of the secondary fragments could be assessed owing to their very different signatures (Figure 1, right insert). The mean dose value based on $\Phi_{prim}$ of 61.45 mGy ± 0.71 % (Figure 1, right, blue line) was 7.4 % lower than that determined by the ionisation chamber of 66.35 mGy ± 0.41 % (green line). According to the transport simulations in Table 2, primary carbon ions account for 97.1 % of the dose, whereas protons (helium) with a relative fluence of 14.8 % (2.4 %) contribute 1.7 % (0.6 %), the influence of heavier fragments is minor. The effective stopping power is therefore very similar to the one from the CSDA approach (Table 1), and taking the energy distribution and secondaries (Table 2) into account reduces the discrepancy $\Delta_C$ by only 2.9 pp leaving 4.5 %.

### Field homogeneity

Forward calculations of the physical beam records have shown that the uniformity of the irradiation fields was within ± 0.8 % for all carbon and proton irradiations. Although the sensitive area of the ionisation chamber is 1.08 cm$^2$ larger than the area of the FNTD, a deviation of 7.4 % between ionisation- and fluence-based dose measurements would mean that the fluence outside of the area covered by both FNTD and ionisation chamber would have been on average 9.6 % higher, which is far beyond the routinely checked constrains of this clinically used system. Further, no significant measured fluence gradients were observed over the length and width of the FNTDs.

### Influence of phantom and FNTD

Small (0.5 % in dose) influence of the RW-3 phantom on the dose to water was seen in the MC simulations in case of the proton beam, mainly due to an increased production of Helium. No similar effect has been seen for the carbon ion beam. The Al$_2$O$_3$ of the FNTD did not change the spectrum significantly within the 20 µm in front of the measurement plane.

## DISCUSSION

Given the uncertainties $\sigma_{TRS}$ as reported in the TRS-398 (2% in $D_{w,Q}$ for p and 3 % for $^{12}$C), for the FNTD (Poisson error, area correction factor), and from experimental design (e.g. inhomogeneous irradiation, machine stability and beam direction), it is believed that the dose assessment of the fluence-based approach agrees with the ionisation-based data in the case of protons. In case of carbon ions, however, the difference is still significant. It is also puzzling that in the carbon beam, the authors detect a relative secondary fluence of approximately $\Phi_{H, He, Li}$ = 3.3 % instead of 17.5 % as predicted by the simulation. If one used these values for dose assessment, the $\Delta_C$ would have been 7.0 % instead of 4.5 %. Even using a more detailed geometrical model of the BAMS including 1 m of air gap to the iso-centre did not reduce $\Phi_{H, He, Li}$ to < 17.0 %.

## CONCLUSION AND OUTLOOK

FNTDs are able to yield correct dose estimation for protons. The assumption of a monoenergetic beam, even in the entrance channel, is invalid since slower protons and secondaries contribute significantly and an effective stopping power has to be employed. These corrections account for the discrepancies seen in the authors' previous experiments. Since the FNTD fluorescent track amplitude depends on the particle species and energy [4]-[5], the effective stopping power might be estimated from the intensity histogram of the particle tracks.

For carbon ions, however, secondary particles did not fully account for the discrepancies found. Considering the detection efficiency of FNTD technology, it seems unlikely that a significant portion of tracks were not registered. This might stimulate discussions on the accuracy of the $k_{Q,Q_0}$ factor for carbon beams [19]. Since the stopping power in this energy range is known quite accurately (1-2 %), one might question the currently used constant $W_{air}$-value of 34.50±0.52 eV (1.5 %) [14]. The presented findings would imply a $W_{air}$-value of 32.10±0.83 eV (2.6 %). This uncertainty includes all conceivable sources of errors including $\sigma_{FNTD}$, $\sigma_{FC}$, and $\sigma_{TRS}$ (except for the uncertainties given for long-term stability of user dosemeter, establishment of reference conditions, dosemeter reading relative to beam monitor and beam quality correction). More conclusive results are expected from absolute dose to water measurements in a carbon ion beam with a water calorimeter, which would allow to directly calibrate ionisation chambers in units of absorbed dose to water without applying radiation-field-dependent correction factors.

ACKNOWLEDGEMENTS

The authors thank Armin Lühr and Katrin Henkner for fruitful discussions and Lorenz Brachtendorf and Naved Chaudhri for their help with the irradiations.

REFERENCES

1. Akselrod, M.S. and Sykora, G.J. Fluorescent nuclear track detector technology - A new way to do passive solid state dosimetry. Radiat. Meas. 46, 1671-1679 (2011).
2. Osinga, J.M., Akselrod, M.S., Herrmann, R., Hable, V., Dollinger, G., Jäkel, O. and Greilich, S. High-accuracy fluence determination in ion beams using fluorescent nuclear track detectors. Radiat. Meas. 56, 294-298 (2013).
3. Akselrod, M.S., Akselrod, A.E., Orlov, S.S., Sanyal, S. and Underwood, T.H. Fluorescent aluminum oxide crystals for volumetric optical data storage and imaging applications. Journal of Fluorescence 13 (6), 503-511 (2003).
4. Sykora, G.J., Akselrod, M.S., Benton, E.R. and Yasuda, N. Spectroscopic properties of novel fluorescent nuclear track detectors for high and low LET charged particles. Radiat. Meas. 43, 422-426 (2008).
5. Niklas, M., Melzig, C., Abdollahi, A., Bartz, J.A., Akselrod, M.S., Debus, J., Jäkel, O. and Greilich, S. Spatial correlation between traversal and cellular response in ion radiotherapy - towards single track spectroscopy. Radiat. Meas. 56, 285-9 (2013).
6. Greilich, S., Osinga, J.M., Niklas, M., Lauer, F.M., Klimpki, G., Bestvater, F., Bartz, J.A., Akselrod, M.S. and Jäkel, O. Fluorescent nuclear track detectors as a tool for ion-beam therapy research. Radiat. Meas. 56, 267-272 (2013).
7. Ambràmoff, M.D., Magalhaes, P.J. and Ram, S.J. Image processing with ImageJ. Biophotonics International 11, 36-42 (2004).
8. Rasband, W.S. ImageJ (version 1.46a). U.S. National Institutes of Health, Bethesda, Maryland, U.S.A. (1997-2011). Available on http://rsbweb.nih.gov/ij/.
9. Cardinale, J. Histogram-based background subtractor for ImageJ. ETH Zurich, Switzerland, (2010).
10. Sbalzarini, I.F. and Koumoutsakos, P. Feature point tracking and trajectory analysis for video imaging in cell biology. Journal of Structural Biology 151, 182-195, (2005).
11. R Development Core Team. R: A Language and Environment for Statistical Computing. R Foundation for Statistical Computing, Vienna (2010). Available on http://www.R-project.org.
12. International commission on Radiation Units and Measurements. Stopping powers and ranges for protons and alpha particles. ICRU Report 49, Bethesda, MD (1993).
13. International commission on Radiation Units and Measurements. Stopping of ions heavier than helium. ICRU Report 73, Bethesda, MD (2005).
14. International atomic energy agency. Absorbed dose determination in external beam radiotherapy. Technical Reports Series No. 398. IAEA, (2000).
15. Jäkel, O., Jacob, C., Schardt, D., Karger, C.P. and Hartmann, G.H. Relation between carbon ion ranges and x-ray CT numbers. Med. Phys. 28, 701-703 (2001).
16. Battistoni, G., Muraro, S., Sala, P.R., Cerutti, F., Ferrari, A., Roesler, S., Fasso`, A. and Ranft, J. The FLUKA code: description and benchmarking. AIP Conference Proceeding 896, 31-49 (2007).
17. Ferrari, A., Sala, P.R., Fasso`, A. and Ranft, J. FLUKA: a multi-particle transport code. CERN-2005-10, INFN/TC_05/11, SLAC-R-773, (2005).
18. Greilich, S., Grzanka, L., Bassler, N., Andersen, C.E. and Jäkel, O. Amorphous track models: a numerical comparison study. Radiat. Meas. 45, 1406-1409 (2010).
19. Hartmann, G.H., Brede, H.J., Fukumara, A., Hecker, O., Hiraoka, T., Jakob, C., Jäkel, O., Krießbach, A. and Schardt, D. Results of a small scale dosimetry comparison with carbon-12 ions at GSI Darmstadt. Proc. Int. Week on Hadrontherapy and 2nd Int. Symp. on Hadrontherapy, Elsevier, 346-350 (1997).